\documentclass[12pt]{article}

\usepackage{scicite}
\usepackage{physics}
\usepackage{times}
\usepackage{pdflscape}
\usepackage{graphicx}
\usepackage{xcolor}
\usepackage{soul}
\usepackage{caption}

\newenvironment{sciabstract}{%
\begin{quote} \bf}
{\end{quote}}

\title{Particles and intrinsic fields supporting topological thermoelectricity}

\author
{Daniel Fa\'ilde,$^{1\ast}$ and Daniel Baldomir$^{1\ast}$\\
\\
\normalsize{$^{1}$Departamento de Física Aplicada, Instituto de Investigacións Tecnolóxicas},\\
\normalsize{Universidade de Santiago de Compostela,}\\
\normalsize{E-15782 Campus Vida s/n, Santiago de Compostela, Spain}\\
\\
\normalsize{daniel.failde.balea@rai.usc.es; daniel.baldomir@usc.es}
}

\date{}

\begin{document} 
\captionsetup[figure]{labelfont={bf},labelformat={default},labelsep=period,name={Fig.}}

\baselineskip24pt

\maketitle

\begin{sciabstract}
At present, topological insulators are the most efficient thermoelectric materials at room temperature. However, at non-zero temperatures, it seems to arise a conflict between having time-reversal symmetry, which implies minimal entropy, and the Seebeck coefficient, which is the entropy carried by each electric charge unit. This has obliged us to analyze the mathematical and physical background taking into account relativistic phonons besides the electrons within quantum field theory. In this search, we found an approximate expression for the intrinsic topological field $b$ in terms of the Chern number,  the Fermi velocity $v_F$ and the electron effective mass $m$, which allows to connect the topologically non-trivial insulator with the trivial one, being consistent with their topological properties and physical robustness. Thanks to this, we demonstrate that for three-dimensional topological insulators in thin-film conditions, among others, phonons have chirality coupling in a novel way to electron dynamics which preserves time-reversal symmetry. This explains the compatibility of the thermoelectricity within topological insulators and shows explicitly how it adapts to the family of topological insulators Bi$_2$Se$_3$.

\end{sciabstract}

Topological insulators (TIs) are Quantum Spin Hall (QSH) systems with conductor states on their surface associated with a non-trivial topology of their electronic band structure \cite{PhysRevLett.96.106802,Konig766}. The presence of spin-orbit coupling (SOC) is one of the most crucial ingredients to get a TI whose role goes beyond its ability to produce a band crossing \cite{zhang2009topological,Xia2009,Bernevig1757}, but it also incorporates an interaction to our system that preserves time-reversal symmetry and gives to electrons a relativistic nature \cite{PhysRevLett.95.226801,PhysRevLett.97.236805}. On the edge of a TI, bands follow a linear dispersion law $E=\hbar v_F k$ as it would happen for a relativistic particle with zero rest mass which goes through singularity points (Dirac points). These singularities are the sources of the non-trivial topology, which, protected by time-reversal symmetry induce chiral Kramers currents that can be determined using Berry's gauge fields associated with their curvature on U(1) or SU(N) groups depending on band degeneracy \cite{RevModPhys.82.3045,PhysRevB.78.195424}. The robustness of these spin-momentum locking channels as well as their relativistic nature, that connects solid state physics with quantum field phenomena, make three and two-dimensional TIs strong candidates in the context of  Quantum Computing or Thermoelectric and Superconducting Devices \cite{leijnse2012introduction,Muchler2013,PhysRevLett.104.057001}. However, for a better understanding of these applications, some key points as thermal excitations and phonons have to be incorporated to electrons dynamic in TIs since their presence could lead to losing the quantum adiabaticity and coherence necessary to maintain such topological order \cite{PhysRevB.81.161302,PhysRevLett.107.210501}.

Along this line, we analyze the possible effect of phonons in TIs surface states by studying mechanical oscillations through Dirac oscillator model \cite{moshinsky1989dirac}, leading to a much richer physical behaviour than the usual employed in the current literature \cite{Mahan}. In order to see if non-trivial topology is preserved under certain values for the involved physical magnitudes in this process, we make use of a field interpretation of the Berry curvature, where magnetic flux quantization of helical currents provides an argument to estimate the strength of the field associated the the topological regime. The applicability of the obtained topological intrinsic field $\boldsymbol{b}$ is determined by the condition $v_F^2>>2MB/\hbar^2$ for the Hamiltonian parameters, which although it seems quite restrictive, it is a good approximation for a wide range of values that typically characterizes two-dimensional (2D) and three-dimensional (3D) TIs \cite{Konig766,PhysRevLett.101.246807,zhang2009topological,PhysRevB.81.115407}. Through this, we give a new interpretation that demonstrates how if a topological phase transition occurrs a spin-dependent field $\boldsymbol{b}$ arises, giving rise to the non-trivial Berry curvature and the quantization of electron transport. In this way, we connect the two vacua that represent the relativistic topological non-trivial regime with the trivial one being consistent with the change on their topological properties at the same time that the magnitude of this field ($\approx10$T for typical parameters $M=-25$meV and $v_F=6.17$ $10^5$m/s in 3DTI thin films) gives an idea of why topological surface states are so robust. Assimilating the previous concepts into the relativistic Dirac equation we found that in its non-relativistic limit the field enters as a spin-orbit coupling term governing Schr\"{o}dinger equation remarking the importance of such interaction in TIs. 

At this point, once we decouple topological signatures from electron wavefunctions we are allowed to compare the effects of oscillations in electrons positions produced by phonons with the internal field within TIs. It is found that both effects are completely equivalent in non-trivial systems with the substitution of  $2\omega=eb/m$ being $\omega$ the phonon frequency, leading to a new heat dissipating mechanism in which electrons can mediate to transform lattice dynamics into electricity when the phonon energies are in the order of the band gap ($2M$). This involves an important result in which the concept of chirality in phonons sources naturally in the previous interaction, which implies no entropy variation in electron-phonon scattering processes through the emission of right-handed and left-handed quanta \cite{PhysRevA.76.041801}. The capacity of TIs to transform heat from the lattice into electricity maintaining quantum coherence  can be one of the main signatures of the topological thermoelectricity, which could explain their high figure of merit \cite{Baldomir2019}. Based on these facts, we focus on a specific scenario in which electron-phonon coupling can be enhanced by interaction with polar optical phonons, as was in numerically calculated and observed experimentally in bulk Bi$_2$Se$_3$, Bi$_2$Te$_3$ and in thin film conditions \cite{Heid2017,PhysRevLett.107.186102,PhysRevB.84.195118}. Given that the the energy $\hbar \omega$ of these polar modes is on the order of the hybridization gap ($\omega\sim$ THz)  for the family of 3DTI Bi$_2$Se$_3$, Bi$_2$Te$_3$ in their thin film limit, we focus on this situation where their top and bottom surfaces hybridize opening a gap on each surface \cite{Shan_2010}, although the results and conclusions obtained are directly applicable to 2DTI as HgTe Quantum Wells (QWs).

\section*{Model}
Chiral edge states can be studied in 3DTI in the same way as for two-dimensional ones, provided by the fact that the thickness $L$ of the material is enough small to overlap the top and bottom surface states. In this limit, in which the system is in a topologically non-trivial regime for certain values of $L$ \cite{PhysRevB.81.115407},  physics is described by a 2D effective Hamiltonian $H_{2D}$ with basis $\left[ \psi_{1\uparrow}, \psi_{2\downarrow}, \psi_{2\uparrow}, \psi_{1\downarrow} \right]$, that can be separated into two non-interacting time-reversal counterparts $H_\pm$, each one with non-zero and opposite Chern numbers \cite{PhysRevB.78.195424,PhysRevB.81.115407,PhysRevB.82.165104}.

\begin{equation}
H_{2D}({\bf k})= \epsilon_0({\bf k})I_{4x4}+\left[\begin{array}{cc} {H_+} & {0} \\
0 & H_- \\
\end{array} \right]
\qquad H_{\pm} = 
\left(\begin{array}{cc} \pm M({\bf k}) & \hbar v_F k_-\\ \hbar v_F k_+ & \mp M({\bf k}) \end{array}\right) \\
\end{equation}
Where $k_\pm=k_x\pm ik_y$, $\epsilon_0({\bf k})=C+Dk^2$, $M({\bf k})=M-Bk^2$ and $k^2=k_x^2+k_y^2$, $v_F$ is the Fermi velocity and $\hbar$ is the Planck's bar constant. To make the notation less confusing,  we have taken the typical form for the off-diagonal terms $v_F \;( \boldsymbol{p} \cdot \boldsymbol{\sigma})$ used in Dirac formalism instead of the $v_F \: (\boldsymbol{p} \times \boldsymbol{\sigma})$ dependence of Bi$_2$Se$_3$ family, this can be done without loss of generality since both terms provide the same Berry curvature and all our calculations remain invariant. For simplicity, we also set to zero the term $\epsilon_0(\textbf{k})$. The energy spectrum define two non-interacting Dirac hyperbolas at the $\Gamma$ point where conduction and valence bands for $H_+$ have associated a Berry curvature of the form
\begin{equation}
    \boldsymbol{\Omega}^{c}_{k_xk_y}=-\boldsymbol{\Omega}^{v}_{k_xk_y}=-\frac{\hbar^2v_F^2(M+Bk^2)}{2[(M-Bk^2)^2+\hbar^2v_F^2k^2]^{3/2}} \boldsymbol{\hat{z}} 
\label{curvature}
\end{equation}
\begin{figure}
    \centering
    \includegraphics[scale=0.50]{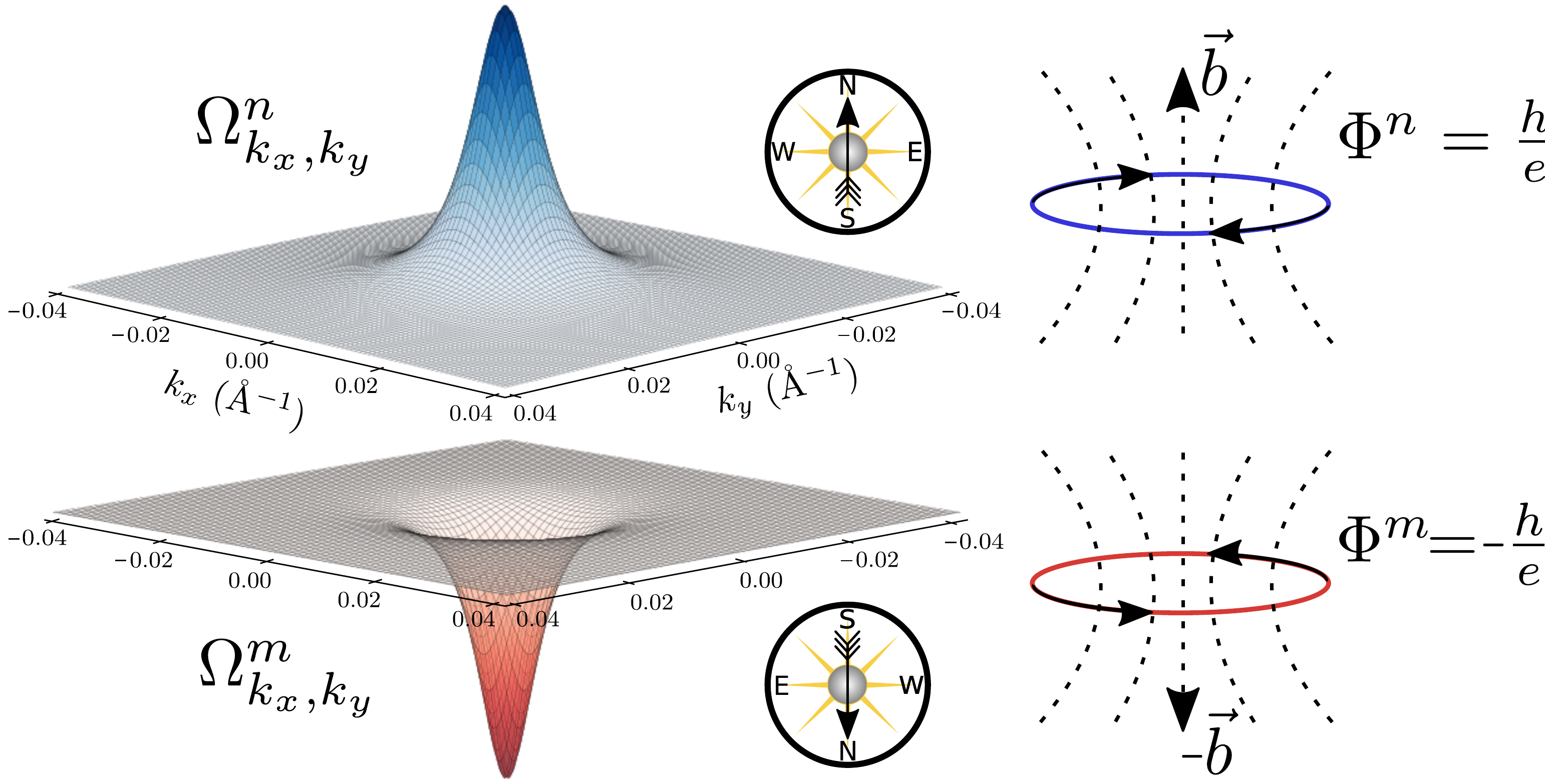}
    \caption{Schematic illustration of the non-trivial ($M<0$, $B<0$) Berry curvature $\boldsymbol{\Omega}_{k_x,k_y}$ for the positive and negative energy eigenstates of $H_+$ labelled as $\ket{n}$ and $\ket{m}$ respectively. The compass indicates the orientation of the field  felt by the electrons on each band. In the bulk of a TI, this curvature allows to consider the existence of helical orbits with an associated flux $\Phi=\hbar/e \; C$, being $C=\pm 1$ the Chern number of the conduction and valence bands of $H_+$ respectively.}
    \label{Fig1}
\end{figure}
that results in opposite sign for $H_-$. This Berry curvature defines in a non-trivial system ($MB>0$) spin-momentum locking orbits with quantized flux ($C \; h/e$) in terms of the Chern number $C$ (Fig.1). That is the reason why in the presence of an in-plane electric field we can talk about opposite transverse spin currents which in the edge produce a quantized electrical conductance $G=(C_+-C_-) e^2/h$, being $C_\pm$ the Chern numbers associated with the branches $H_\pm$ \cite{Konig766}. In terms of transport, this is one of the main signatures that differentiate the bulk of a trivial insulator from a non-trivial one. Both phenomena the quantization of the magnetic flux and the conductance are completely consistent with presence of a spin-dependent $\textbf{b}$ field in TIs that must be the source of the non-trivial Berry curvature. Under small gap conditions, as it happens for 3DTI surface states in the thin film limit, we can estimate the magnitude of this field by noticing that the non-trivial Berry curvature $\boldsymbol{\Omega}_{k_x,k_y}$ has a form of a single peak Gaussian-like function centered at the $\Gamma$ point \cite{PhysRevB.81.115407}, which has a characteristic length small enough to consider an equivalent magnetic field $\boldsymbol{b}$ constant along the bulk crystal and whose magnitude must be determined by the constraint that his flux must be quantized and equal to $h/e \; C$
\begin{figure}
    \centering
    \includegraphics[scale=0.39]{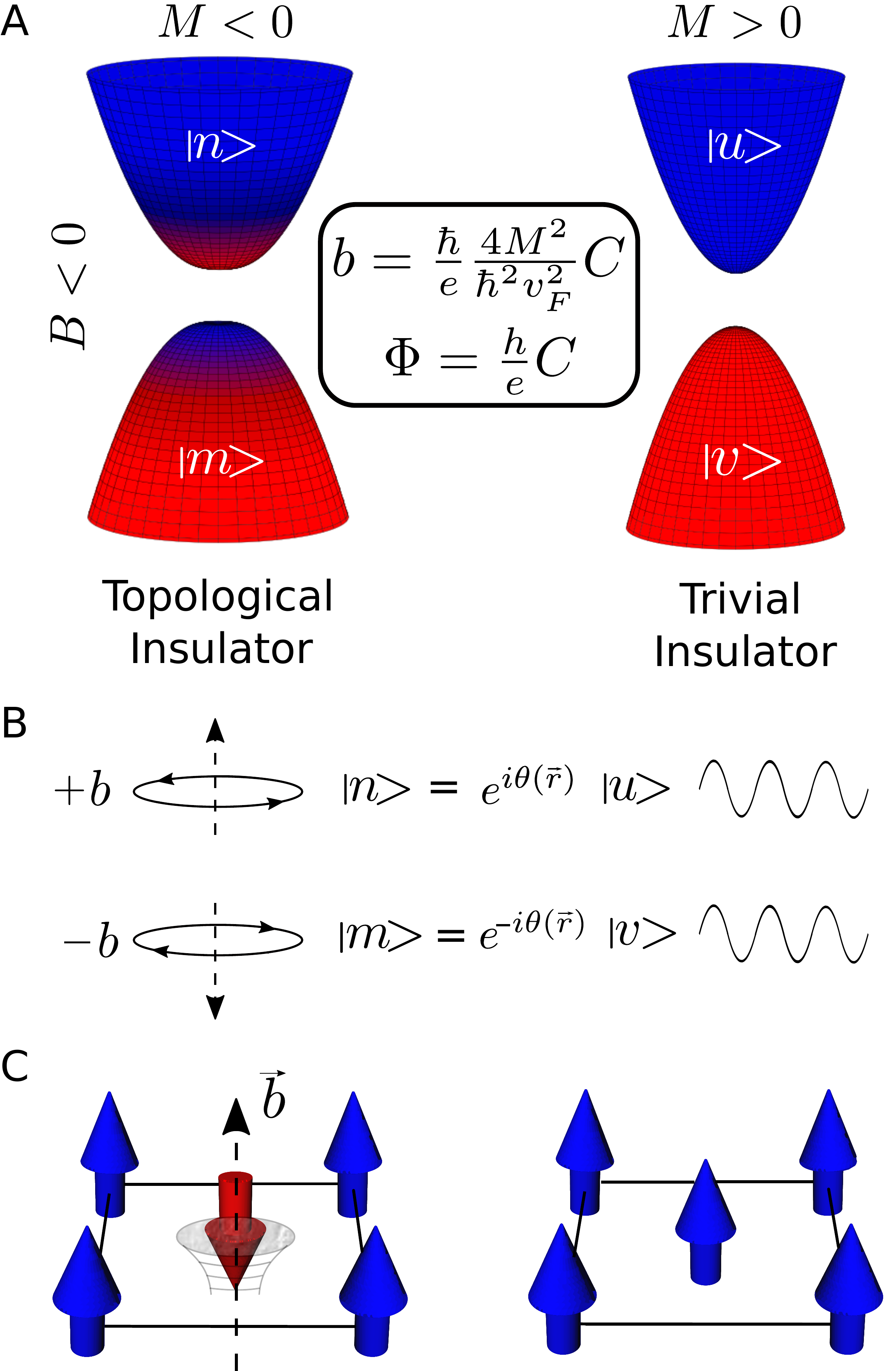}
    \caption{\textbf{(A)} Energy spectrum of the 2x2 Dirac Hamiltonian $H_+$ in both non-trivial (left) and trivial (right) topological regimes. Conduction and valence bands are labelled with their respective positive and negative energy eigenstates. Red and blue colors stand for electrons spin attending to the different spin configurations of both regimes. The change in topological properties between both systems allows us to define the intrinsic topological field $\boldsymbol{b}$ in terms of the Chern number $C$. This expression is exact for systems with $B=0$ and a good approximation for systems whose parameters satisfy the condition $v_F^2>>2MB/\hbar^2$. \textbf{(B)} Connection between the eigenstates of the non-trivial and trivial topological regimes in terms of a magnetic phase factor $\theta(\boldsymbol{r})$ coming from the field $\boldsymbol{b}$. This field connects the free-particle behaviour in the trivial regime with the quantized electronic transport in TIs. \textbf{(C)} Schematic illustration of spin configuration in the k-space for the conduction band of $H_+$. The existence of the field $\boldsymbol{b}$ in the topological non-trivial regime is consistent with the change of electron spin at the $\Gamma$ point.}
    \label{Fig2}
\end{figure} 
 \begin{equation}
     \frac{\hbar}{e} \int \boldsymbol{\Omega}_{k_x,k_y} d\boldsymbol{k} = \int \boldsymbol{b} d\boldsymbol{S}
 \end{equation}
The surface element $\Delta \boldsymbol{S}$ can be obtained in a fashion way by applying Heisenberg's uncertainty principle matching the quantum conductance ($e^2/h$) with the conductivity $\sigma=\Delta S^{-1} \frac{e^2 \tau}{m_{eff}}$ in the Heisenberg limit ($\tau = \hbar/2\Delta E$). In this way, we obtain the following expression for the field
 
 \begin{equation}
    \textbf{b}=\frac{\hbar}{e} \frac{4M^2}{\hbar^2 v_F^2} C \; \hat{\textbf{z}}
\label{field}
\end{equation}
where we have considered the electron effective mass as $m_{eff}=M/v_F^2$ (Appendix I), neglecting any contribution from $B$, that gives us the information about the localization or delocalization of the bands in the space, and limiting to materials that present $v_F^2>>2MB/\hbar^2$. This condition, easily fulfilled thanks to the small gap and high Fermi velocity that typical characterizes 3DTI thin films ($M\approx -25$ meV, $v_F=6.17 \; 10^5$m/s), determines for these values an equivalent field for electrons on the surface $b\approx10$T consistent with the robustness that characterizes topological surface states and whose sign, determined by the Chern number, keeps time-reversal symmetry intact. For the relativistic system with $B=0$ the prior condition is always satisfied and then Eq. \ref{field} is accurate, although by definition the integral of its Berry curvature results in a half-integrer number. Even then, in both cases ($B=0$ or $B\neq0$) the variation between the two different topological regimes, let us say the trivial ($M>0$ or $MB<0$) and the non-trivial ($M<0$ or $MB>0$), leads to the same magnitude for $b$, determining the field that connects one regime with the other.

This field will be crucial to puzzle out electron dynamics in TIs (Fig.2). On the one hand, we have that in a TI electrons move in closed helical orbits, with an associated quantized flux to their Berry curvature that acts as an spin-dependent magnetic field and whose integral gives a non-zero Chern number. On the other hand, a relativistic trivial insulator ($MB<0$) has zero Chern number and their curvature has no quantized flux associated, obtaining in its non-relativistic limit the Schr\"odinger equation for a free electron. The crucial step is that these two pictures are connected exactly by the field $\boldsymbol{b}$

\begin{equation}
    \ket{n}=e^{-i\frac{e}{\hbar}\int \boldsymbol{a_n} d\boldsymbol{r}} \ket{u}
\label{transformation}
\end{equation}
being $\ket{u}$ the eigenstate with positive energy of a trivial insulator, i.e $H_N\ket{u}=\xi_N\ket{u}$, $\ket{n}$ the eigenstate with positive energy of $H_{TI}$ ($MB>0$), and $\boldsymbol{a}_n=(-by/2,bx/2,0)$ the vector potential associated to a field $\boldsymbol{b}$ which contains the information of Berry curvature effects and allow us to pass from topologically trivial Berry curvature to a non-trivial one and vice versa (Fig.3). This is straightforward to see by computing the Berry curvature, obtaining that $\Omega_{k_x,k_y}^u=\Omega_{k_x,k_y}^n+2\Omega_{k_x,k_y}^{\bar{m}}$, being $\ket{\bar{m}}=\ket{m(B=0)}$ the eigenstate with an associated field of exactly $\boldsymbol{-b}/2$. Integrating over the 2D surface one gets the expected result
\begin{figure}[ht]
    \centering
    \includegraphics[scale=0.17]{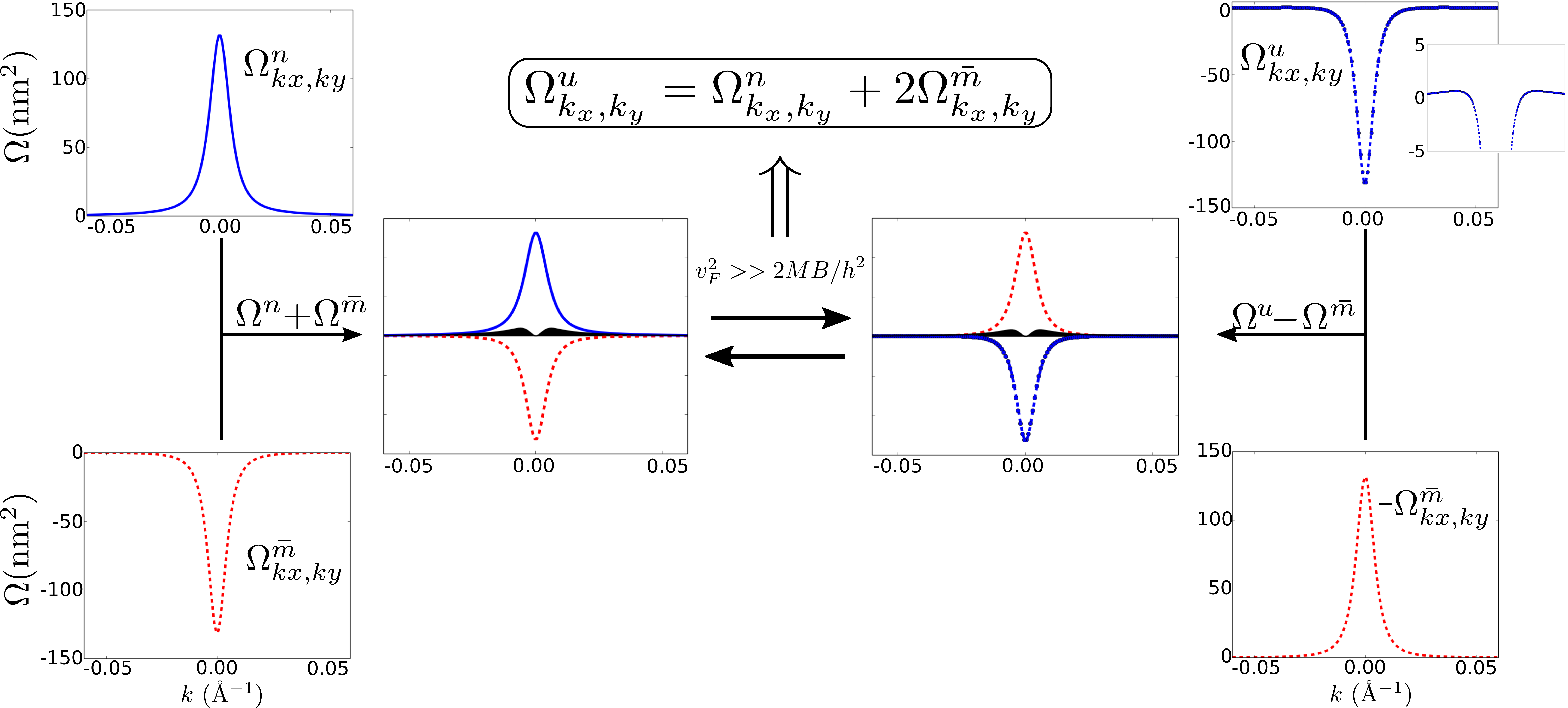}
    \caption{Conduction band Berry curvature ($B\neq 0$) as a function of $k$ for non-trivial (solid blue line) and trivial (dotted-dashed blue line) topological regimes. Red dotted lines correspond to the Berry curvature for the relativistic case $B=0$ where $\Omega^{\bar{m}}_{k_x,k_y}$ has an associated field of exactly $-b/2$. Central plots illustrate the superposition of $\Omega^{n}_{k_x,k_y}+\Omega^{\bar{m}}_{k_x,k_y}$ and $\Omega^{u}_{k_x,k_y}-\Omega^{\bar{m}}_{k_x,k_y}$ (blak areas), whose result is equivalent when the condition $v_F^2>>2MB/\hbar^2$ is fulfilled, allowing to pass from one topological regime to the other. Plot parameters are $M=-0.025$eV, $B=-50$eV\AA$^2$ and $v_F=6.17$ $10^5$m/s while $B=0$ for $\Omega^{\bar{m}}_{k_x,k_y}$ curves.}
    \label{Fig3}
\end{figure}
\begin{equation}
\int \boldsymbol{\Omega}_{k_x,k_y}^u d\boldsymbol{k}= \int \boldsymbol{\Omega}_{k_x,k_y}^n d\boldsymbol{k}+\int 2\boldsymbol{\Omega}_{k_x,k_y}^{\bar{m}} d\boldsymbol{k} = 2\pi\left(1-2\frac{1}{2}\right)=0
\end{equation}
Consistent with this, extending the previous calculations to $H_-$ and rearranging the basis to $\left[ \psi_{1\uparrow}, \psi_{1\downarrow}, \psi_{2\uparrow}, \psi_{2\downarrow} \right]$ we can establish a relationship between both trivial and non-trivial regimes by writing the Dirac equation $i\hbar\partial_t \Psi_{TI}=H_{TI}\Psi_{TI}$, where applying Eq. \ref{transformation} for the positive energy solution one gets  
\begin{equation}
\xi_{TI} \Psi_N=[M({\bf k})\beta +v_F\boldsymbol{\alpha}(\boldsymbol{p}-ieb\boldsymbol{r}/2\beta)]\Psi_N \; , \qquad  \Psi_N=\left[ \begin{array}{c} \psi_{1,N} \\ \psi_{2,N} \end{array} \right]
\label{TopolicalHamiltonian}
\end{equation}
$\psi_{1,N}$ and $\psi_{2,N}$  are two component spinors for electrons in the trivial regime where the subindex 1/2 stands for the different atoms that conforms our 2D effective Hamiltonian. Diagonal corrections are negligible due to the fact that they are proportional to $2MB/\hbar^2v_F^2<<1$. Notice that this connection can be done provided that the sign of the field is opposite between the branches of the Hamiltonian $H_\pm$ and thus, there is no time-reversal symmetry breaking as it occurs in TIs. Therefore, in compliance with the conventional treatment of TIs by means of Berry potentials and curvature, we give a field interpretation to understand topological phase transitions, in the sense that once the crossing between conduction and valence bands occurs a spin-dependent field appears in the system, opening the gap and giving an effective mass to the particles likewise it happens in other quantum phase transitions. Following up this line, we can  directly compute the non-relativistic limit of Eq. \ref{TopolicalHamiltonian},

\begin{equation}
    \epsilon \psi_{1/2,N}= \left(\frac{p^2}{2m}+\frac{e^2b^2r^2}{8m} \mp \mu_B b \mp \frac{eb}{m \hbar} L_z S_z \right) \psi_{1/2,N}
\end{equation}
where $m$ is the electron effective mass, $\mu_B$ is the Bohr magneton, $L_z$ and $S_z$ are the z-components of the angular momentum operator and the spin operator respectively. This limit allows us to show explicitly how an electron in a trivial insulator sees its topological counterpart. On the one hand, we have that the term $\mu_B b$, whose magnitude is the band gap $2M$, contains the information of the band crossing necessary to pass from one system to the other. On the other hand, the second and fourth term act as a magnetic field with opposite sign between spin and between solutions, defining a spin-orbit coupling term for the global system, whose magnitude is defined as one outcome which translates the abstract topological robustness, usually used in the literature without being explained its real associated strength and its possibility of measurement in the laboratory. Furthermore, as we are going to see, by means of this field interpretation of non-trivial effects, we also obtain a picture in which phonons can couple in a novel way to electrons, giving a theoretical framework that fits well in the mechanism of the thermoelectricity and superconductivity.

\section*{Electron-phonon interaction in TIs}
The introduction of mechanical oscillations in a relativistic context was first analyzed by M. Moshisnky and A. Szczepaniak \cite{moshinsky1989dirac} incorporating a linear term in $r$ to the Dirac equation. The origin of this term lies in the introduction of an harmonic oscillator potential into the Klein-Gordon equation, leading to the well-known Dirac oscillator
\begin{equation}
i\hbar (\partial \psi / \partial t)=[v_F \boldsymbol{\alpha}(\boldsymbol{p}-im \boldsymbol{r}\omega\beta)+mv_F^2\beta] \psi
\label{Dirac}
\end{equation}
being $\alpha_i=\left[\begin{array}{cc} {0} & \sigma_i \\
\sigma_i & 0 \\
\end{array} \right]$, $\beta=\left[\begin{array}{cc} \sigma_0 & 0 \\
0 & -\sigma_0 \\
\end{array} \right]$ , $\sigma_i$ the Pauli matrices, $m$ the mass of the particle, $r$ the position and where we have substituted the original speed of light $c$ by the Fermi velocity $v_F$ in order to adapt Eq. \ref{Dirac} into the context of TIs. Afterwards, the equation was further analyzed, always in the context of Quantum Field Theory (QFT) \cite{Rozmej_1999, PhysRevA.76.041801}, where  working with phonons it is always convenient to employ operators defined on a Fock space and Eq. \ref{Dirac} has a simple solution in 2D for them \cite{PhysRevA.76.041801}. There are two Pauli spinor eigenstates which present entanglement between the spin and orbital degrees of freedom. That is to say, they superpose positive and negative energies with respect to their vacuum state due to have a much lower relativistic energy by being associated to the Fermi velocity instead of the Dirac's one employing light velocity. Notice that the Stone-Von Neumann theorem \cite{Hall}, only follows in quantum mechanics with finite degrees of freedom, but not in QFT which justifies different vacua states seen by the operators within the TI material. In the Dirac vacuum the phonon could never excite electrons or be absorbed by them, among other reasons because phonons do not exist in it, while in TIs they can do it provided the conservation of angular momenta and energy are followed and their vacua are connected (Appendix II). In most materials, the phonons are complex collective excitation arising with different frequencies, polarizations and moving in chaotic trajectories, but fortunately in TIs or in 2D materials they can be controlled enough well to measure their special characteristics \cite{PhysRevLett.107.186102,PhysRevLett.98.166802,PhysRevB.88.064307,PhysRevLett.108.187001}.

These abstract remarks are fundamental to understand the phonon-electron interaction within the TIs. In fact, if one compare Eq. \ref{TopolicalHamiltonian} and Eq. \ref{Dirac} one can easily see that they are essentially the same and both terms enters in the same way in electrons relativistic dynamics with the usual substitution $2\omega=eb/m$ with a difference of a factor 2 that comes from the non-minimal coupling in the Dirac oscillator equation that guaranties to have an harmonic oscillator in its non-relativistic limit with a spin-orbit coupling of strength $2\omega/\hbar$ \cite{moshinsky1989dirac,Andrade_2014}

\begin{equation}
    \epsilon \psi= \left(\frac{p^2}{2m}+\frac{1}{2} m \omega^2 r^2 \mp \hbar \omega \mp \frac{2 \omega}{\hbar} L_z S_z \right) \psi
\end{equation}
\begin{figure}
    \centering
    \includegraphics[scale=0.6]{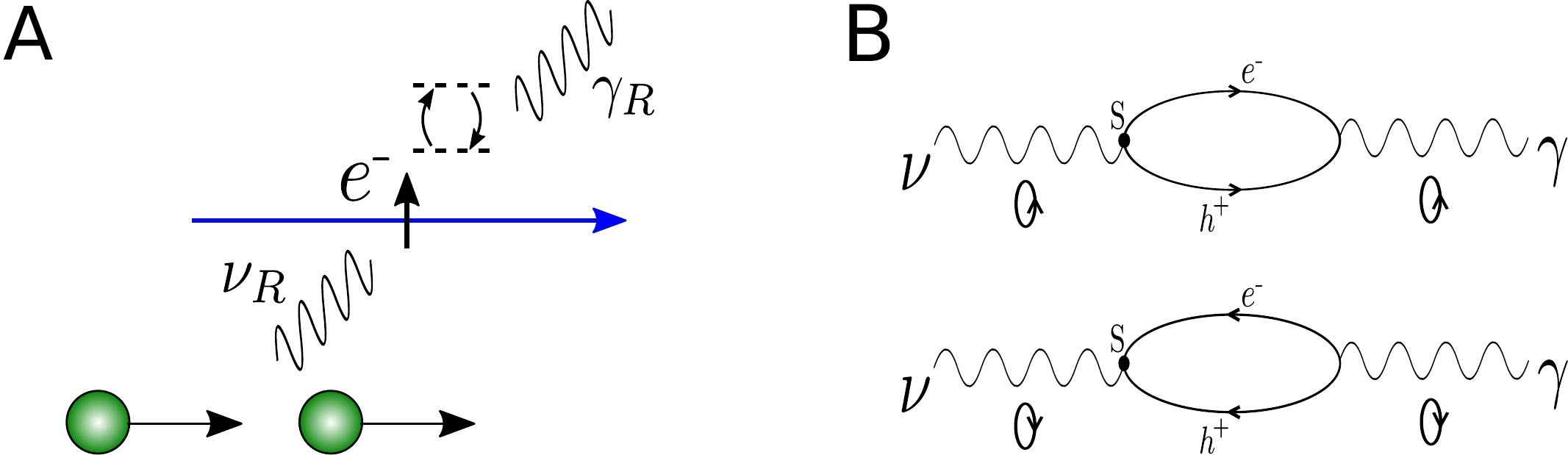}
    \caption{\textbf{(A)} Schematic illustration of electron-phonon interaction between right-handed electrons and phonons producing the emission of  definite chiral photons preserving angular momenta. Green circles represent the nucleus of the lattice \textbf{(B)} Feynman diagrams associated to the electron-phonon coupling in TIs. The point S stands from the Schwinger’s limit, i.e, when the energy is enough to create an electron-hole pair.}
    \label{Fig4}
\end{figure}
We know from Eq. \ref{TopolicalHamiltonian} that the electrons in a TI follow a spin-orbit interaction for their spins and now we see that phonons couple to them in the same way. Provided by the spin polarization that defines the non-trivial topology, manifested by the existence of chiral edge states with spin-orbit locking channels, the conservation of angular momenta as a good quantum number is possible in a same Hamiltonian allowing the electrons to be excited by the phonons as if they were photons. In other words, we found that in a topologically non-trivial system, electrons are capable to dissipate heat from the lattice through photons thanks to the polarized electron currents that originate the  intrinsic topological field (Fig.4). This is consistent with the preservation of time-reversal symmetry at non-zero temperatures \cite{PhysRevLett.107.210501}, where electronic interactions could in principle break it if they change the entropy of the system, helping to maintain the quantum coherence necessary to preserve topological properties at high temperatures \cite{Park2015,Zhao2014}. We can write Eq. \ref{Dirac} in terms of right-handed and left-handed chiral annihilation and creation operators \cite{PhysRevA.76.041801}, to describe an scenario where the chirality of phonons have a decisive role in mechanism of thermoelectricity in TI . In fact, it has been shown \cite{Heid2017} that for the 3DTIs Bi$_2$Se$_3$ and Bi$_2$Te$_3$ the polar optical phonon modes satisfy the conditions $2\omega=eb/m$ underlined in this article, being accompanied with an enhance of the electron-phonon coupling when the Fermi level lies close to the Dirac point. Under this conditions, we provide the ingredients necessary to minimize lattice thermal conductivity in TIs, leading to an experimental scenario in which we can get a topological thermoelectric system with a maximized figure of merit. \cite{Baldomir2019,Venkatasubramanian2001,Xu2017}.

\section*{Discussion}
Topological insulators are materials which have a semimetal behavior on their surfaces due to the crossing band singularities. At low energies, this allows to interpret the dispersion law of the electrons relativistically, when the rest mass of the electrons is given to be zero and, it is well known, how this relativistic background works very well in the literature. But, beside that, it is assumed a time-reversal symmetry in half-integer spin systems, which induces Kramers degenerated currents without backscattering. They are justified just by the non-trivial topology calculated through Gauss integrals of the Berry curvature without never mentioning the microscopic physical structure needed to support it. This is the main achievement of this paper, we have found that the Dirac singularities must be accompanied by a quite high magnetic field $b$ associated directly to the Berry field and its Chern number. In spite of its high value, around Teslas, it is very difficult to measure directly since its sign is opposite between states resulting in the form of a spin-orbit term. Nevertheless, it is easy to see that this magnetic field is consequence of the field created by Berry's phase around the band singularities, that we have found in different forms and which directly supports the robust topological structure of TIs. Non-trivial topology of one manifold can exist only when the manifold is mantained rigid enough to keep it. In other words, we have shown that this physical representation is consistent with the topological properties of TIs, quantized transport, robustness, $k$-dependent spin-configuration and band inversion, describing an scenario in which it is straightforward to connect both trivial and non-trivial as we demonstrate expressing the field $b$ in terms of the Chern number. We also shown the applicability of the model given by the condition $v_F^2>> 2MB/\hbar^2$, which is fulfilled for the family of 3DTIs Bi$_2$Se$_3$ as well as in 2D for HgTe Quantum Wells (QWs).
On the other hand, we know that these materials have an excellent thermoelectric response and therefore phonons must follow also the relativistic context of the electrons, in such a form that their exchange of energy and momentum enables adiabaticity on the bands. This has obliged us to treat the phonons also within a quantum field Fock space instead of the usual usual quantum mechanical formalism. The main result was to obtain also chiral phonons as it ought to be if the electron-phonon coupling were able to behave enough adiabatically smoothly. Actually, this allows the phonons to be absorved by the electrons as the topological thermoelectricity would need to survive.  
Finally, we calculate the non-relativistic limit of the Dirac Hamiltonian, for electrons and phonons, where the spin-orbit coupling arises naturally thanks to employing the relativistic formalism and the von Neumann entropy associated to the states can be kept practically constant preserving the time-reversal symmetry at least on the surface of these materials. In summary, the topological structure is kept robust under thermal changes due to the electron-phonon coupling and the potentials of deffects are not enough large to change the trajectories under the high field strength associated with the band singularities. 

\bibliography{Particles_and_intrinsic_fields}

\begin{thebibliography}{10}

\bibitem{PhysRevLett.96.106802}
B.~A. Bernevig, S.-C. Zhang, {\it Phys. Rev. Lett.\/} {\bf 96}, 106802 (2006).

\bibitem{Konig766}
M.~K{\"o}nig, {\it et~al.\/}, {\it Science\/} {\bf 318}, 766 (2007).

\bibitem{zhang2009topological}
H.~Zhang, {\it et~al.\/}, {\it Nature physics\/} {\bf 5}, 438 (2009).

\bibitem{Xia2009}
Y.~Xia, {\it et~al.\/}, {\it Nature Physics\/} {\bf 5}, 398 (2009).

\bibitem{Bernevig1757}
B.~A. Bernevig, T.~L. Hughes, S.-C. Zhang, {\it Science\/} {\bf 314}, 1757
  (2006).

\bibitem{PhysRevLett.95.226801}
C.~L. Kane, E.~J. Mele, {\it Phys. Rev. Lett.\/} {\bf 95}, 226801 (2005).

\bibitem{PhysRevLett.97.236805}
S.~Murakami, {\it Phys. Rev. Lett.\/} {\bf 97}, 236805 (2006).

\bibitem{RevModPhys.82.3045}
M.~Z. Hasan, C.~L. Kane, {\it Rev. Mod. Phys.\/} {\bf 82}, 3045 (2010).

\bibitem{PhysRevB.78.195424}
X.-L. Qi, T.~L. Hughes, S.-C. Zhang, {\it Phys. Rev. B\/} {\bf 78}, 195424
  (2008).

\bibitem{leijnse2012introduction}
M.~Leijnse, K.~Flensberg, {\it Semiconductor Science and Technology\/} {\bf
  27}, 124003 (2012).

\bibitem{Muchler2013}
L.~M{\"u}chler, F.~Casper, B.~Yan, S.~Chadov, C.~Felser, {\it Phys. Status
  Solidi RRL\/} {\bf 7}, 91 (2013).

\bibitem{PhysRevLett.104.057001}
Y.~S. Hor, {\it et~al.\/}, {\it Phys. Rev. Lett.\/} {\bf 104}, 057001 (2010).

\bibitem{PhysRevB.81.161302}
R.~Takahashi, S.~Murakami, {\it Phys. Rev. B\/} {\bf 81}, 161302 (2010).

\bibitem{PhysRevLett.107.210501}
M.~B. Hastings, {\it Phys. Rev. Lett.\/} {\bf 107}, 210501 (2011).

\bibitem{moshinsky1989dirac}
M.~Moshinsky, A.~Szczepaniak, {\it Journal of Physics A: Mathematical and
  General\/} {\bf 22}, L817 (1989).

\bibitem{Mahan}
G.~D. Mahan, {\it Many-Particle Physics\/} (Springer, Boston, MA, 2000).

\bibitem{PhysRevLett.101.246807}
B.~Zhou, H.-Z. Lu, R.-L. Chu, S.-Q. Shen, Q.~Niu, {\it Phys. Rev. Lett.\/} {\bf
  101}, 246807 (2008).

\bibitem{PhysRevB.81.115407}
H.-Z. Lu, W.-Y. Shan, W.~Yao, Q.~Niu, S.-Q. Shen, {\it Phys. Rev. B\/} {\bf
  81}, 115407 (2010).

\bibitem{PhysRevA.76.041801}
A.~Bermudez, M.~A. Martin-Delgado, E.~Solano, {\it Phys. Rev. A\/} {\bf 76},
  041801 (2007).

\bibitem{Baldomir2019}
D.~Baldomir, D.~Fa{\'i}lde, {\it Scientific Reports\/} {\bf 9}, 6324 (2019).

\bibitem{Heid2017}
R.~Heid, I.~Y. Sklyadneva, E.~V. Chulkov, {\it Scientific Reports\/} {\bf 7},
  1095 (2017).

\bibitem{PhysRevLett.107.186102}
X.~Zhu, {\it et~al.\/}, {\it Phys. Rev. Lett.\/} {\bf 107}, 186102 (2011).

\bibitem{PhysRevB.84.195118}
V.~Gnezdilov, {\it et~al.\/}, {\it Phys. Rev. B\/} {\bf 84}, 195118 (2011).

\bibitem{Shan_2010}
W.-Y. Shan, H.-Z. Lu, S.-Q. Shen, {\it New Journal of Physics\/} {\bf 12},
  043048 (2010).

\bibitem{PhysRevB.82.165104}
H.~Li, L.~Sheng, D.~N. Sheng, D.~Y. Xing, {\it Phys. Rev. B\/} {\bf 82}, 165104
  (2010).

\bibitem{Rozmej_1999}
P.~Rozmej, R.~Arvieu, {\it Journal of Physics A: Mathematical and General\/}
  {\bf 32}, 5367 (1999).

\bibitem{Hall}
H.~B.C, {\it Quantum Theory for Mathematicians\/} (Springer, New York, 2013).

\bibitem{PhysRevLett.98.166802}
J.~Yan, Y.~Zhang, P.~Kim, A.~Pinczuk, {\it Phys. Rev. Lett.\/} {\bf 98}, 166802
  (2007).

\bibitem{PhysRevB.88.064307}
Y.~Wang, L.~Guo, X.~Xu, J.~Pierce, R.~Venkatasubramanian, {\it Phys. Rev. B\/}
  {\bf 88}, 064307 (2013).

\bibitem{PhysRevLett.108.187001}
Z.-H. Pan, {\it et~al.\/}, {\it Phys. Rev. Lett.\/} {\bf 108}, 187001 (2012).

\bibitem{Andrade_2014}
F.~M. Andrade, E.~O. Silva, {\it {EPL} (Europhysics Letters)\/} {\bf 108},
  30003 (2014).

\bibitem{Park2015}
B.~C. Park, {\it et~al.\/}, {\it Nature Communications\/} {\bf 6}, 6552 (2015).

\bibitem{Zhao2014}
L.~Zhao, {\it et~al.\/}, {\it Nature Materials\/} {\bf 13}, 580 (2014).

\bibitem{Venkatasubramanian2001}
R.~Venkatasubramanian, E.~Siivola, T.~Colpitts, B.~O'Quinn, {\it Nature\/} {\bf
  413}, 597 (2001).

\bibitem{Xu2017}
N.~Xu, Y.~Xu, J.~Zhu, {\it npj Quantum Materials\/} {\bf 2}, 51 (2017).

\bibitem{shen2012topological}
S.-Q. Shen, {\it Topological insulators\/}, vol. 174 (Springer, 2012).

\bibitem{Batra2002}
I.~P. Batra, {\it Solid State Communications\/} {\bf 124}, 463 (2002).

\bibitem{Umezawa}
H.~Umezawa, {\it Advanced Field Theory: Micro, Macro and Thermal Physics\/}
  (American Institute of Physics, NY, 1995).

\end{thebibliography}

\bibliographystyle{Science}

\section*{Acknowledgements}
\textbf{Funding:} Authors acknowledge to CESGA, AEMAT ED431E 2018/08 and the MAT2016-80762-R project for financial support. \textbf{Author contributions:} D.F. and D.B. conceived the problem, made the calculations and wrote the manuscript. \textbf{Competing Interests:} The authors declare no competing interests.

\newpage

\section*{Supplementary Materials}

\subsection*{Appendix I}

It is well-known that low energy physics close to Dirac points can be modelled through an effective Dirac Hamiltonian which depends on the system dimensionality. In 2D and in 3DTI thin films, electrons dynamics are described by a Hamiltonian of the form \cite{PhysRevB.81.115407,Shan_2010,PhysRevLett.101.246807} 

\begin{equation}
H({\bf k})= \epsilon_0({\bf k})I_{4x4}+\left[\begin{array}{cccc} {M({\bf k})} & {\hbar v_Fk_-} & {0} & {0} \\
{\hbar v_Fk_+} & {-M({\bf k})} & {0} & {0} \\
{0} & {0} &  {-M({\bf k})} & {\hbar v_Fk_-} \\
{0} & {0} & {\hbar v_Fk_+} & {M({\bf k})} 
\end{array} \right]
\label{Original Hamiltonian}
\end{equation}
where $k_\pm=k_x\pm ik_y$, $\epsilon_0({\bf k})=C+Dk^2$, $M({\bf k})=M-Bk^2$ and $k^2=k_x^2+k_y^2$. Due to the absence of $k_z$, the previous Hamiltonian is separated into two independent subsystems $H_\pm$ that are time-reversal counterparts 

\begin{equation}
H_{\pm} = 
\left(\begin{array}{cc} \pm M({\bf k}) & \hbar v_F k_-\\ \hbar v_F k_+ & \mp M({\bf k}) \end{array}\right)
\end{equation} 
In this frame, working with one of these Hamiltonian (in our case $H_+$) the topological analysis can be made in terms of the Chern number instead of the usual $Z_2$ formalism for time-reversal symmetric systems analyzing the parity changes at the time-reversal invariant points \cite{shen2012topological}. Notice, that we have chosen a explicit form for the off-diagonal terms proportional to $v_F \; \boldsymbol{p} \cdot \boldsymbol{\sigma}$ although there also exists the possibility to have a term of the form $v_F \; \boldsymbol{p} \times \boldsymbol{\sigma}$ \cite{Shan_2010}. It is easy to prove that both expressions leads to the same Berry curvature making our election irrelevant in order to calculate the field $\boldsymbol{b}$. Thanks to this separation we have a direct quatization of electrons transport in terms of the Chern number $C=1/2\pi \int \boldsymbol{\Omega}_{k_x,k_y} d^2\boldsymbol{k}$ which is zero for a trivial system ($MB<0$) and an integer in a non-trivial one ($MB>0$). This basic number derived from the curvature of the electrons wavefunction in the $\boldsymbol{k}$-space takes into account their singular transport properties, i.e, quantized electric conductance ($e^2/h \; C$) at the semi-metallic edges and their bulk-correspondent helical orbits with quantized magnetic flux $\Phi=h/e \; C$. From this picture, that clearly determines the differences between the topologically non-trivial and trivial regimes, it naturally sources the presence of a field $\boldsymbol{b}$ that gives rise to the non-trivial Berry curvature once the topologically phase transition ocurrs

\begin{equation}
    \frac{\hbar}{e} \int \boldsymbol{\Omega} \; d\boldsymbol{k}=\int  \boldsymbol{b} \; d\boldsymbol{S}
\end{equation}
The concept of single electrons moving in quantized orbits allow us to apply the Heisenberg uncertainty principle $\tau \Delta E \sim \hbar/2$ where the energy uncertainty can be supposed to be on the order of the energy $E$ provided by the low-energy nature of the electrons \cite{Batra2002}. Thus, to be consistent with the fact that we are establishing a relationship between the non-trivial regime and the trivial one we define $\Delta E=2mv_F^2$, that is, the energy difference between both regimes. Now, we can compute directly the surface element $\Delta S$ by matching the quantum conductance ($e^2/h$) to the conductivity $\sigma=(\Delta S)^{-1} \frac{e^2\tau}{m}$ obtaining 

\begin{equation}
    \Delta S =\frac{h \hbar}{4m^2v_F^2}
\end{equation}
Once obtained the surface element we can give an expression to the topological intrinsic field extracting it from the integral. This can be done, provided that for a wide range of values under small gap conditions the Berry curvature is a narrow single peak function centered at the $\Gamma$ point and thus we can consider an equivalent field constant along the crystal. Furthermore, it is also easy to compute an expression for the electron effective mass $m$ on each band

\begin{equation}
m=\frac{\hbar^2}{\partial^2E/\partial k^2} \qquad
 \qquad   E_\pm=\pm \sqrt{(M-Bk^2)^2+\hbar^2v_F^2k^2}
\end{equation}
which in the limit $k\rightarrow0$ gives

\begin{equation}
m=\frac{M}{v_F^2-2BM/\hbar^2}= \left\{ \begin{matrix} M/v_F^2 & if \quad v_F^2>>2BM/\hbar^2
\\ -\hbar^2/2B & if \quad v_F^2<<2BM/\hbar^2 \end{matrix} \qquad  \right.
\end{equation}
The first condition, in which we obtain an effective mass equal to half the Dirac gap, it is fullfilled by the Hamiltonian parameters in 2DTI and 3DTI thin films, giving an accurate expression to the topological intrinsic field 

\begin{equation}
    b=\frac{\hbar}{e} \frac{4M}{\hbar^2 v_F^2} C
\end{equation}
that it is exact for the traditional Dirac Hamiltonian with $B=0$.

\pagestyle{empty}

\subsection*{Appendix II}

The electrons of the topological insulators are modelled within a quantum relativistic context and also seems natural to do the same with the phonons interacting with them. In fact, their physics corresponds to a Quantum Field Theory (QFT) better than a Quantum Mechanical (QM) system and they turn out to be bosons with a well-defined chirality. Treating them as spinors, it is convenient to do it within a Fock space whose creation and annihilation operators appears also with chirality, albeit their origin is not on Kramers theorem as the electrons. But their vacua do not need to be equivalent under unitary transformations, because in QFT there are infinite degrees of freedom at difference of what happens in QM which follows the Stone-von Neumann theorem. Nevertheless, we shall show that, despite to have a non-trivial topology, their Fock spaces can transform their vacua thanks to introduce the new Berry field $b$, which is an important property for their thermoelectric properties.

The Dirac oscillator equation 
\begin{equation}
i\hbar (\partial \psi / \partial t)=[v_F \boldsymbol{\alpha}(\boldsymbol{ p }-im \boldsymbol{ r }\omega\beta)+mv_F^2\beta] \psi
\end{equation}
can be rewritten in function of the chiral annhilation and creation operators $a_r = \frac{1}{ \sqrt{2}}(a_x -ia_y), a_r^+ = \frac{1}{\sqrt{2}}(a_x^+ +ia_y^+)$ for the right and $a_l = \frac{1}{\sqrt{2}}(a_x +ia_y), a_l^+ = \frac{1}{\sqrt{2}}(a_x^+ -ia_y^+)$ for the left, being $a_x, a_y,a_x^+$ and $a_y^+$ the usual annhilation and creation operators of the harmonic oscillator
\begin{equation}
    \ket{\psi_1}=i\frac{2mv_F^2\sqrt{\xi}}{E-mv_F^2}a_l^+\ket{\psi_2} 
\end{equation}

\begin{equation}
    \ket{\psi_2}=-i\frac{2mv_F^2\sqrt{\xi}}{E+mv_F^2}a_l^+\ket{\psi_1}
\end{equation}
being $\ket{\psi_1}$ and $\ket{\psi_2}$ the two components of the spinor $\ket{\psi}$, $\xi=\frac{\hbar \omega}{m v_F^2}$ takes into account the non-relativistic limit and the Fock space is expanded by the basis $\ket{n_l}= \frac{1}{\sqrt{n_l!}}(a_l^+)^{n_l}\ket{0}$ \cite{PhysRevA.76.041801}. The energy spectrum is $E= \pm E_{n_l}= \pm m v_F^2 \sqrt{4 \xi{n_l +1}}$, whose eigenstates can be written as Pauli spinors $\ket{\phi_\uparrow}$ and $\ket{\phi_\downarrow}$ components, employing the angular momentum z-component definition given by  $L_z=\hbar (a_r^+a_r-a_l^+a_l)$

\begin{equation} 
    \ket{-E_{n_l}}=\beta_{n_l}\ket{n_l}\ket{\phi_\uparrow}+i\alpha_{n_l}\ket{n_l-1}\ket{\phi_\downarrow}
\end{equation}

\begin{equation}
    \ket{E_{n_l}}=\alpha_{n_l}\ket{n_l}\ket{\phi_\uparrow}-i\beta_{n_l}\ket{n_l-1}\ket{\phi_\downarrow}
\end{equation}
where $\alpha_{n_l}=\sqrt{\frac{E_{n_l}+mv_F^2}{2E_{n_l}}}$ and $\beta_{n_l}=\sqrt{\frac{E_{n_l}-mv_F^2}{2E_{n_l}}}$. Finally, time dependent state of the spinors excited by the Dirac oscillator is
\begin{equation}
    \ket{\psi(t)}=\left(cos \omega_{n_l}t+\frac{i}{\sqrt{4\xi n_l+1}}sin \omega_{n_l}t   \right) \ket{n_l-1} \ket{\phi_\uparrow} +\left(\sqrt{\frac{4\xi_{n_l}}{4\xi_{n_l}+1}} sin \omega_{n_l}t\right)\ket{n_l} \ket{\phi_\downarrow}
\end{equation}
Therefore, we see how there is one oscillation between the spin-orbit states $|n_l-1> |\phi_\uparrow>$ and $|n_l> |\phi_\downarrow>$,in such a form that the change of spin polarization implies one for the orbital and vice versa.  At the same time this is done on the surface of the topological insulator where there is a simultaneous change between the conduction and valence bands, i.e. positive and negative energies respect to the Fermi level.

The fact that  the relativistic phonons have chirality like the electrons does produce a condensed phase formed by their pairs within the Thermal Field Dynamics (TFT). Let us show it \cite {Umezawa} assuming that the phonon $a, a^+$ and photon $ b, b^+$, annilation and creation operators, repectively, have the only non vanishing commutation relation $[a_g,a_f^+]=[b_g,b_f^+]= \delta^2(g-f)$. Introducing the unitary transformations $U=exp(i \theta G) $, being $G=i(ab-b^+a^+)$ the generator with $\theta$ the c-number associated to the boson translation $\alpha(\theta)=a+\theta$ and $\beta(\theta)=b+\theta$. This translation obviously keeps the commutation of the operators invariant, or the algebras, while the vacuum doesn't do it. We can relate them by the Bogoliubov transformations
\begin{equation}
\alpha(\theta)=U(\theta)\; a\; U^{-1}(\theta)=a \; cosh (\theta) -b^+ sinh( \theta)
\end{equation}
\begin{equation}
\beta(\theta)=U(\theta)\; a\; U^{-1}(\theta)=b \; cosh (\theta) -a^+ sinh( \theta)
\end{equation}
where $\alpha(\theta)$ and $\beta(\theta)$ follow again the same commutation rules that $a$ and $b$ operators although their vacua are different,  $\alpha(\theta)\ket{\theta}=\beta(\theta)\ket{\theta}=0$ while $a\ket{0}=b\ket{0}=0$, i.e. being $\alpha(\theta)\ket{0}=\beta(\theta)\ket{0}=\theta(0)$, which is different to zero in the Fock space $H(a,b)$. Thus their associated Fock spaces are different because they have not the same vacua. But they are related by
\begin{equation}
\ket{0(\theta)}=exp \left( -ln\; cosh(\theta) \right) exp \left( a^+b^+tanh(\theta)\right) \ket{0}
\end{equation}
whose projection is given by
\begin{equation}
\bra{0}\ket{0(\theta)}=exp\left(-ln\; cosh(\theta)\right) 
\end{equation}
Now taking into account that we are within QFT instead of QM, the operators and vacuum depend also of the wave number $k$ of the modes where they work with momentum $p=\hbar k$ and therefore that the above formulae must be added for these different modes,i.e.
\begin{equation}
\bra{0}\ket{0(\theta)}= exp[-\delta^2 (0) \int  ln\; cosh(\theta_k) \; d^2k]\;  exp[\int tanh (\theta_k)\; a^+_k b^+_k \; d^2k)]
\end{equation}
which is zero because the integral is infinite just noticing that $\delta^2(0)=\delta(k_x)\delta(k_y)|_{k_x=k_y=0}=\infty$ where k is with only two components. This means that the Fock spaces associated to these vacua are orthogonal besides to be different and therefore we cannot transform their tensors using unitary groups. That is to say, the Stone-von Neumann the QM cannot be employed in this QFT physical context. But it is worthy to observe that we have avoided this singular point introducing a Berry curvature on the point $k=0$ or its associated field $b$. The physical interpretation is that we have pairs $a,b$ associated to phonon-photon pairs at a given temperature $T$ related by a chiral electron. The phonon excite the charge of the electron which produces a photon carrying the same same chirality for conserving the angular momentum: under thermal changes both chiralities can be keeped invariant.And the thermoelectric properties of the topological insulators lie under this basic property because the process phonon-electron-photon is not so probable as the photon-electron-phonon in the condensation of the phase for these particles in the $\ket{0(\theta)}$ vacuum.

\end{document}